\documentclass[a4paper,10pt]{article}
\usepackage{amsfonts}
\usepackage{amsmath}
\usepackage{amssymb}
\usepackage{graphicx}

\begin{document}

\title{Ballistic transport properties in pristine/doped/pristine graphene
junctions}
\author{J.S. Ardenghi$^{\dag }$\thanks{%
email:\ jsardenghi@gmail.com, fax number:\ +54-291-4595142}, P. Bechthold$%
^{\dag }$, E. Gonzalez$^{\dag }$, P. Jasen$^{\dag }$ and A. Juan$^{\dag }$ \\
$^{\dag }$IFISUR, Departamento de F\'{\i}sica (UNS-CONICET)\\
Avenida Alem 1253, Bah\'{\i}a Blanca, Argentina}
\maketitle

\begin{abstract}
We investigate the ballistic electron transport in a monolayer graphene with
configurational averaged impurities, located between two clean graphene
leads. It is shown that the electron transmission are strongly dependent on
the concentration of impurities and the incident energy. In turn, the
conductance computed using the Landauer formalism shows a similar behavior
to those found in experimental works as a function of the applied voltage
for different concentrations of impurities in the limit of low temperatures.
In the limit of zero bias voltage, the conductance shows a minimum value
which reduces to zero for high concentration of impurites which disentangle
graphene sublattices. These results can be very helpful for exploring the
tunneling mechanism of electrons through doped thermodynamically stable
graphene.
\end{abstract}

\section{Introduction}

Graphene, a new material with promising application possibilities and
important fundamental physics aspects, is a two-dimensional allotrope of
carbon which has become one of the most significant topics in solid state
physics (\cite{novo},\cite{intro1},\cite{intro2}, \cite{B}, \cite{BBBB}).
The carbon atoms form a honey-comb lattice made of two interpenetrating
triangular sublattices, $A$ and $B$. A special feature of the graphene band
structure is the linear dispersion at the Dirac points which are dictated by
the $\pi $ and $\pi ^{\prime }$ bands that form conical valleys touching at
the high symmetry points of the Brillouin zone \cite{A}. Electrons near
these symmetry points behave as massless relativistic Dirac fermions with an
effective Dirac-Weyl Hamiltonian \cite{B}.

The understanding of charge transport mechanism of pristine graphene/doped
graphene is crucial for future applications in nanoelectronics. As clean
graphene being a gapless semimetal, is useless for electronic development,
therefore it is necessary to turn graphene from semimetallic to a gap
semiconductor, which can be realized in several ways (with substrates \cite%
{sh} by confinement \cite{han} and quantum dots \cite{pono}). In the other
side, a quasigap in the vicinity of the Dirac point can be obtained in two
dimensional system lattice sites with two different site energies and
different probabilities \cite{yu}. In this work we report a theoretical
model to describe the transport mechanism in the ballistic regime at the
interface of clean graphene and graphene with adatoms with arbitrary
energies (positive values as donor and negative values as acceptor) placed
on a site-like position. The impurities are randomized and averaged over its
possible positions which transform the diffusive system in a ballistic one.
In this sense, the system can be considered in thermodynamic equilibrium
with a fixed number of impurity concentration. The diffusive character of
systems with disorder can be study through Green function techniques (see 
\cite{janis}, \cite{zie}, \cite{rammer}) and Kubo formalism, which allow to
obtain the quantum corrections to the conductivity and other effects as weak
antilocalization.\footnote{%
Although for ideal graphene, the dynamics of the electrons produce the same
shot noise as that found in classical diffusion (see \cite{mario}).} For the
purpose of this work, we will consider that the quantum mechanical coherence
length is longer than the sample size $L$, in this case, the disorder of
random impurities is transformed in a mass term in the Hamiltonian by the
averaging procedure. Then by applying Landauer formalism (\cite{landa}) is
enough to study the ballistic behavior of Bloch electrons through the sample.%
\footnote{%
Other theoretical methods can be applied to obtain the effects of impurities
in the electronic spectra of graphene (see \cite{feher}).} Graphene-based
devices have not been fully investigated due to the complex processes
required to achieve p- and n-doped semiconducting graphene. By chemical
doping, graphene-based p-n junctions can be obtained (\cite{will}, \cite{lo}%
), but graphene retain its semimetallic character. In this work, from the
impurity averaged tight-binding Hamiltonian in the long wavelength limit, is
possible to obtain a Dirac equation with mass for the Bloch electrons. In
this sense, a gap in the energy band near the Fermi energy is obtained. The
tunnel junction studied is based on pristine graphene/doped
graphene/pristine graphene (PG/DG/PG). The transport mechanism is highly
dependent on the impurity concentrations and several predictions can be
obtained through theoretical calculations to be applied to different
metal/semiconductor junctions (see \cite{bai}).

The paper is organized as follows. In Section II, we introduce the
tight-binding Hamiltonian with impurities and the averaging procedure. Then,
the long-wavelength approximation is applied to obtain Dirac equation with
mass. In Section III, Landauer formalism to the PG/DG/PG system is applied
for low temperatures. The conductance as a function of the applied voltage
is obtained. Minimum conductance is computed for zero voltage. Finally, an
equation relating the length of the sample as a function of the
concentration of impurities can be computed to obtain a transmission
coefficient which is indepenent on the impurity concentration.

In the final section, the conclusions are presented. In appendix A, the
coefficients of the Taylor expansion of the conductance as a function of the
impurities are computed.

\section{Tight binding model with impurities}

The tight-binding Hamiltonian of graphene for nearest neighbors reads%
\begin{equation}
H_{0}=-t\underset{\left\langle i,j\right\rangle ,\sigma }{\overset{}{\sum }}%
(a_{i,\sigma }^{\dag }b_{j,\sigma }+b_{i,\sigma }^{\dag }a_{j,\sigma })
\label{1}
\end{equation}%
where $a_{i,\sigma }^{\dag }$($a_{i,\sigma }$) creates (annihilates) an
electron on site $\mathbf{r_{i}}$ with spin $\sigma $, where $\sigma
=\uparrow ,\downarrow $ on sublattice $A$ and $b_{i,\sigma }^{\dag }$($%
b_{i,\sigma }$) creates (annihilates)\ an electron on site $\mathbf{r_{i}}$
with spin $\sigma $, on sublattice $B$ and $t$ is the nearest neighbor $%
\left\langle i,j\right\rangle $ hopping energy. Impurities can be included
in the tight-binding description by the addition of a local energy term%
\begin{equation}
H_{imp}=\underset{i,\sigma }{\overset{N_{i}}{\sum }}(V_{i}a_{i,\sigma
}^{\dag }a_{i,\sigma }+W_{i}b_{i+\delta ,\sigma }^{\dag }b_{i+\delta ,\sigma
})  \label{2}
\end{equation}%
where $V_{i}$ is a random potential at site $\mathbf{r_{i}}$ and $W_{i}$ is
a random potential at site $\mathbf{r_{j}}$ and where $\mathbf{\delta }%
=a(1,0,0)$. By introducing the Fourier transform of the annihilation and
creation operators $a_{i,\sigma }$ and $b_{i,\sigma }$: 
\begin{equation}
a_{i,\sigma }=\frac{1}{\sqrt{N}}\underset{\mathbf{k}}{\overset{}{\sum }}e^{i%
\mathbf{k\mathbf{r_{i}}}}a_{\mathbf{k},\sigma }\text{ \ \ \ \ \ \ }%
b_{i,\sigma }=\frac{1}{\sqrt{N}}\underset{\mathbf{k}}{\overset{}{\sum }}e^{i%
\mathbf{k\mathbf{r_{i}}}}b_{\mathbf{k},\sigma }  \label{3a}
\end{equation}%
The Hamiltonian reads%
\begin{equation}
H=\underset{\mathbf{k,}\sigma }{\overset{}{\sum }}\left[ \phi (\mathbf{k)}a_{%
\mathbf{k},\sigma }^{\dag }b_{\mathbf{k},\sigma }+\phi ^{\ast }(\mathbf{k)}%
b_{\mathbf{k},\sigma }^{\dag }a_{\mathbf{k},\sigma }\right] +\frac{1}{N}%
\underset{\mathbf{k,\mathbf{q,}}\sigma }{\overset{}{\sum }}\left( V(\mathbf{q%
})a_{\mathbf{k},\sigma }^{\dag }a_{\mathbf{k+\mathbf{q}},\sigma }+W(\mathbf{q%
})b_{\mathbf{k},\sigma }^{\dag }b_{\mathbf{k+\mathbf{q}},\sigma }\right)
\label{4a}
\end{equation}%
where%
\begin{equation}
V(\mathbf{q})=\underset{i}{\overset{N_{i}}{\sum }}V_{i}e^{i\mathbf{qr_{i}}}%
\text{ \ \ \ \ \ \ \ \ }W(\mathbf{q})=\underset{i}{\overset{N_{i}}{\sum }}%
W_{i}e^{i\mathbf{qr_{i}}}  \label{5a}
\end{equation}%
A configurational averaging over the impurites can be applied over the
Hamiltonian of last equation, that is, the lattice points where the
impurities are located can be placed at random with different random
configurities.\footnote{%
In general, configurational averaging is applied over the Green function
(see \cite{rammer}). The average restore the translation symmetry of the
system, but transforms the original system of non-interacting electrons to a
correlated one. Configurational averaging applied directly over the
Hamiltonian restore traslation invariance and the impurities appears as an
effective mass term. In this sense, the average over the Hamiltonian
disables the disorder introduced in eq.(\ref{4a}).} We can sum over all the
configurations of possible positions of impurities in the lattice. If there
are $N_{i}$ impurities, then the configurational averaging can be computed
as (see \cite{rammer})%
\begin{equation}
\left\langle F\right\rangle =\frac{1}{A^{N_{i}}}\underset{\mathbf{r_{1}}}{%
\overset{}{\sum }}...\underset{\mathbf{r_{N_{i}}}}{\overset{}{\sum }}F(%
\mathbf{r_{1},...,r_{N_{i}}})  \label{6a}
\end{equation}%
where $A$ is the area of graphene sheet. The configurational averaged
Hamiltonian reads%
\begin{equation}
\left\langle H\right\rangle =\underset{\mathbf{k}}{\overset{}{\sum }}\left[
\phi (\mathbf{k)}a_{\mathbf{k},\sigma }^{\dag }b_{\mathbf{k},\sigma }+\phi
^{\ast }(\mathbf{k)}b_{\mathbf{k},\sigma }^{\dag }a_{\mathbf{k},\sigma }+Va_{%
\mathbf{k},\sigma }^{\dag }a_{\mathbf{k},\sigma }+Wb_{\mathbf{k},\sigma
}^{\dag }b_{\mathbf{k},\sigma }\right]  \label{6.1a}
\end{equation}%
where%
\begin{equation}
V=\frac{1}{N}\underset{i}{\overset{N_{i}}{\sum }}V_{i}\text{ \ \ \ \ \ }W=%
\frac{1}{N}\underset{i}{\overset{N_{i}}{\sum }}W_{i}  \label{7a}
\end{equation}%
and%
\begin{equation}
\left\vert \phi (\mathbf{k})\right\vert =-t\sqrt{1+4\cos ^{2}(\frac{\sqrt{3}%
}{2}k_{y}a)+4\cos (\frac{3}{2}k_{x}a)\cos (\frac{\sqrt{3}}{2}k_{y}a)}
\label{7.1a}
\end{equation}%
and where we have used eq.(\ref{6a}). Hamiltonian of eq.(\ref{6.1a}) can be
diagonalized and the spectrum reads%
\begin{equation}
E_{\lambda }(\mathbf{k})=\frac{V+W}{2}+\lambda \sqrt{(\frac{V-W}{2}%
)^{2}+\left\vert \phi (\mathbf{k})\right\vert ^{2}}  \label{9a}
\end{equation}%
where $\lambda =\pm 1$, where the plus sign is for the conduction band and
the minus sign for the valence band. To study the behavior of electrons at
the Dirac point, we can expand the energy near the $K$ Dirac point$(\frac{%
2\pi }{3a},\frac{2\pi }{3\sqrt{3}a})$%
\begin{equation}
E_{\lambda }(\mathbf{p})=\lambda \sqrt{\frac{\gamma ^{2}}{4}+\frac{3}{4}%
\frac{t^{2}a^{2}}{\hbar ^{2}}p^{2}}  \label{10a}
\end{equation}%
where $\gamma =V-W$ and $\mathbf{p=\hbar k}$ and where we the constant term $%
\frac{V+W}{2}$ has been absorbed into a redefinition of the energy. This
low-energy description is valid as long as the characteristic energy is
smaller than a cutoff $E_{C}\sim \frac{\hbar v_{F}}{a}\sim 2.6eV$ of the
order of the inverse lattice spacing (see \cite{peres-guinea}).

The last Hamiltonian is the Hamiltonian of a massive Dirac fermions, where
the mass and velocity reads%
\begin{equation}
m=\frac{2\gamma \hbar ^{2}}{3t^{2}a^{2}}  \label{11a}
\end{equation}%
and%
\begin{equation}
v_{F}=\frac{\sqrt{3}ta}{2\hbar }  \label{12a}
\end{equation}%
which is identical to the Fermi velocity of clean graphene. In this sense,
graphene with impurities in the long wavelength approximation can be
considered as fermions satisfying the Dirac Hamiltonian with mass given by
eq.(\ref{11a}) and velocity given by eq.(\ref{12a}). The mass of electrons
is proportional to $\gamma $, which depends on the impurity potentials. In
the case that $V_{i}=V_{0}$ and $W_{i}=W_{0}$, $V=n_{i}V_{0}$ and $W=$ $%
n_{i}W_{0}$, where $n_{i}=N_{i}/N$ is the impurity concentration. The mass
term due to the impurities can be interpreted as if graphene is altered by a
periodic potential which originates an effective mass for the propagation of
electrons. This mass term is not a diffusive term in the Hamiltonian, then
it cannot gives information about many-body effects for electrons in
graphene with impurities. Nevertheless, impurities are still there in the
Hamiltonian as a mass term and can give a detailed description for ballistic
transport phenomena, provided that the quantum mechanical coherence length
is longer than the sample size $L$. Even more, graphene with random
impurities is no longer a disorder system when the average is applied to the
Hamiltonian, but inertial effects appears in electrons.

\subsection{Dirac equation}

By apply the quantization procedure to the energy of eq.(\ref{9a}) we can
obtain the Dirac equation%
\begin{equation}
E\psi =(v_{F}\overrightarrow{\sigma }\cdot \overrightarrow{p}+\sigma
_{z}mv_{F}^{2})\psi  \label{d1}
\end{equation}%
where $\overrightarrow{\sigma }$ are the Pauli matrices $\sigma _{x}$ and $%
\sigma _{y}$. The solution of the Dirac equation reads%
\begin{equation}
\psi _{\lambda }(\mathbf{r})=\frac{1}{\sqrt{2}}\left[ 
\begin{array}{c}
\sqrt{1+\frac{\lambda mv_{F}^{2}}{\sqrt{m^{2}v_{F}^{4}+v_{F}^{2}p^{2}}}} \\ 
\lambda e^{i\phi _{p}}\sqrt{1-\frac{\lambda mv_{F}^{2}}{\sqrt{%
m^{2}v_{F}^{4}+v_{F}^{2}p^{2}}}}%
\end{array}%
\right] e^{\frac{i}{\hbar }\mathbf{p\cdot r}}  \label{d2}
\end{equation}%
where 
\begin{equation}
\phi _{p}=arctg(\frac{p_{y}}{p_{x}})  \label{d3}
\end{equation}%
In the limit of no impurities,~$m=0$ the spinor of eq.(\ref{d2}) is
identical to the one of clean graphene as it is expected.\footnote{%
The limit $m=0$ can be obtained with $\gamma =0$ which implies that both
impurity potentials $V$ and $W$ can be identical.} In the limit of high
concentration of impurities $m\rightarrow \infty $ the spinor decouples the
pseudospin%
\begin{equation}
\underset{m\rightarrow \infty }{\lim }\psi _{+}(r)=\left[ 
\begin{array}{c}
1 \\ 
0%
\end{array}%
\right] e^{\frac{i}{\hbar }\mathbf{p\cdot r}}  \label{tb20}
\end{equation}%
and 
\begin{equation}
\underset{m\rightarrow \infty }{\lim }\psi _{-}(r)=\left[ 
\begin{array}{c}
0 \\ 
1%
\end{array}%
\right] e^{\frac{i}{\hbar }\mathbf{p\cdot r}}  \label{tb21}
\end{equation}%
which implies that the impurities break the symmetry of $A$ and $B$
sublattice.

As we said before, graphene with random impurities has been transformed to
graphene with a periodic potential that introduces an effective mass which
is proportional to the impurity concentration. In this sense, the wave
function of eq.(\ref{d2}) encodes the probability amplitude of finding a
Bloch electron in the whole sample of lentgh $L$. This wave function do not
contains information about a possible localization of the electron due to
the interference effects of the impurities. Nevertheless, if the sample of
length $L~$is placed near a sample of clean graphene, the wave functions
must match in the boundary, which introduces several restrictions to the
energies involved for the electron propagation. The fact that electrons in
graphene with a configurational averaging of impurities applied to the
Hamiltonian introduces an effective mass proportional to the impurity
concentration, implies that the restrictions to the energies involved will
be sensitive to the impurities through the mass term. From this result, it
is possible to obtain a detailed description of the transport phenomena in
the ballistic regime where the wave function is not localized.

\section{Tunneling transport through a potential barrier}

Consider the following two-dimensional model where doped graphene is placed
between two reservoirs with different chemical potentials $\mu _{L}$ and $%
\mu _{R}$ connected through pristine graphene leads (see Figure 1). By
matching the solution of Dirac equation for $x<0$ with $m=0$ with the
solution of Dirac equation for $0<x<L$ (see eq.(\ref{d2})) and with the
solution of Dirac equation for $x>L$ we obtain for the transmission as a
function of the incident energy\footnote{%
The $y$ direction has a length $D$ which introduces a quantization of the $%
p_{y}$ component of the momentum.}%
\begin{equation}
T_{n}(E)=\frac{16\left\vert \eta _{n}\right\vert ^{2}}{\left\vert e^{-i\xi
_{n}L}(\eta _{n}+1)^{2}-e^{i\xi _{n}L}(\eta _{n}-1)^{2}\right\vert ^{2}}
\label{tt1}
\end{equation}%
where 
\begin{equation}
\eta _{n}(E)=\frac{\hbar v_{F}(\xi _{n}-i\alpha _{n})}{E-E_{0}}  \label{tt2}
\end{equation}%
and%
\begin{equation}
\xi _{n}(E)=\frac{1}{\hbar v_{F}}\sqrt{E^{2}-E_{0}^{2}-\hbar
^{2}v_{F}^{2}\alpha _{n}^{2}}  \label{tt3}
\end{equation}%
where $E_{0}=mv_{F}^{2}$ and $\alpha _{n}=\frac{n\pi }{D}$, where $D$ is the
width of the sample ribbon. The allowed energy are in the range $%
-E_{C}<E<E_{C}$. For the $n=0$ mode, the transmission coefficient reads%
\begin{equation}
T_{0}(E)=\frac{16\eta _{0}^{2}}{16\eta _{0}^{2}+4(\eta _{0}^{2}-1)^{2}\sin
^{2}(\xi _{0}L)}  \label{tt3.8}
\end{equation}%
\begin{figure}[tbp]
\centering
\includegraphics[width=110mm,height=50mm]{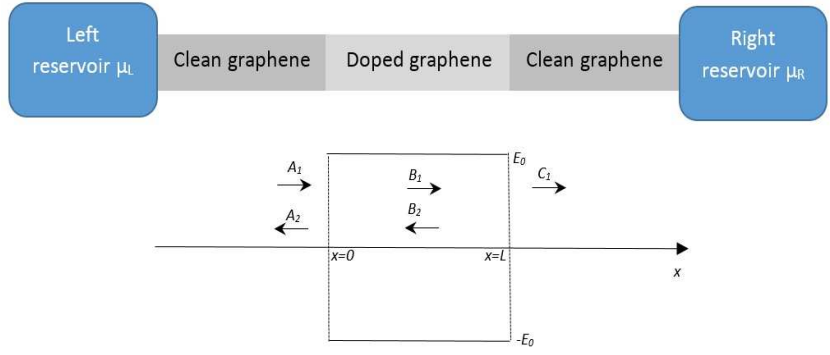}
\caption{Schematic illustration of the tunnel junction composed of pristine
graphene in the left and right leads and graphene with impurities in the
scattering region.}
\label{schematic}
\end{figure}
where the maximum are located when $\xi _{0}L=m\pi $, for $m=\in 
\mathbb{Z}
$%
\begin{equation}
E_{\max }=\sqrt{\left( \frac{\hbar v_{F}m\pi }{L}\right) ^{2}+E_{0}^{2}}
\label{tt3.9}
\end{equation}%
The first mode $n=0$ is independent of the width of the ribbon $D$ due to
the the factor $\alpha _{n}=\frac{n\pi }{D}$, which implies that the
conductance and transmission are valid for any value of $D$.\footnote{%
This could be appropiate for some experimental purpose, since it is not
necessary to consider one of the sample dimension.} The Landauer formula for
the total current flowing from the left to right lead in the $n=0$ mode is
given by%
\begin{equation}
I_{T}=I_{L\rightarrow R}-I_{R\rightarrow L}=g_{s}\frac{2\left\vert
e\right\vert }{\hbar }\int_{-E_{C}}^{E_{C}}dET_{0}(E)(f_{L}(E,\mu
_{L})-f_{R}(E,\mu _{R}))  \label{tt4}
\end{equation}%
where $g_{s}=2$ is the spin degeneracy factor and where $f(E,\mu )$ is the
Fermi-Dirac distribution%
\begin{equation}
f_{L(R)}(E,\mu _{L(R)})=\frac{1}{1+\exp [\beta (E-\mu _{L(R)})]}  \label{tt5}
\end{equation}%
where $\beta =1/k_{B}T$. If we assume that the chemical potentials are
related as $\mu _{R}=\mu _{L}+\left\vert e\right\vert V_{SD}$ where $V_{SD}$
is the source-drain voltage difference between the leads and we shift the
energy as $E\rightarrow E+\frac{eV_{SD}}{2}$, then the difference of Fermi
distributions can be written as a gate function%
\begin{equation}
g(\beta )=f_{L}(E,\mu _{L})-f_{R}(E,\mu _{R})=-\frac{\sinh (\frac{\beta
eV_{SD}}{2})}{\cosh (\frac{\beta eV_{SD}}{2})+\cosh (\beta (E-\mu _{L}))}
\label{tt6}
\end{equation}%
\begin{figure}[tbp]
\centering
\includegraphics[width=95mm,height=55mm]{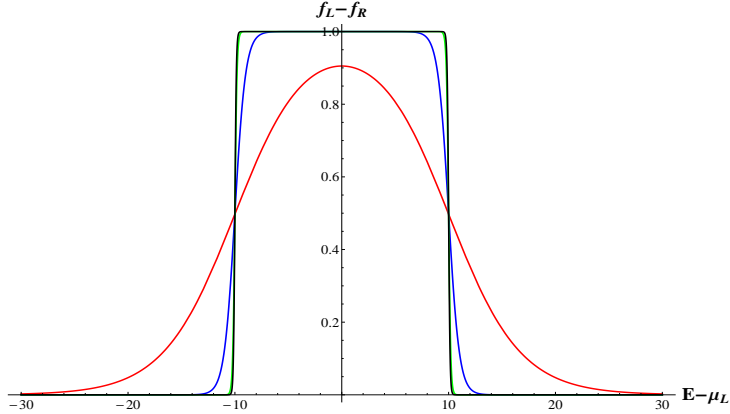}
\caption{Gate function for different temperatures (red line $\protect\beta %
=0.3$, blue line $\protect\beta =2$, green line $\protect\beta =10$ and
black line $\protect\beta =15$, $\left\vert e\right\vert V_{SD}=20$). }
\label{gate}
\end{figure}
which behaves at low temperatures as $f_{L}(E,\mu _{L})-f_{R}(E,\mu
_{R})\sim 1\,$between $\mu _{L}-\frac{eV_{SD}}{2}$ and $\mu _{L}+\frac{%
eV_{SD}}{2}$ and zero in the remaining energy values (see \ref{gate}). As $%
\beta $ goes down, the gate function goes to zero by relaxing the behavior
of rectangular function. For $\beta \rightarrow \infty $, the integral of
eq.(\ref{tt4}) reads%
\begin{equation}
\int_{-E_{C}}^{E_{C}}dET_{0}(E)(f_{L}(E,\mu _{L})-f_{R}(E,\mu
_{R}))=\int_{\mu _{L}}^{\mu _{L}+eV_{SD}}T_{0}(x)dx  \label{au20}
\end{equation}%
\begin{figure}[tbp]
\centering
\includegraphics[width=95mm,height=55mm]{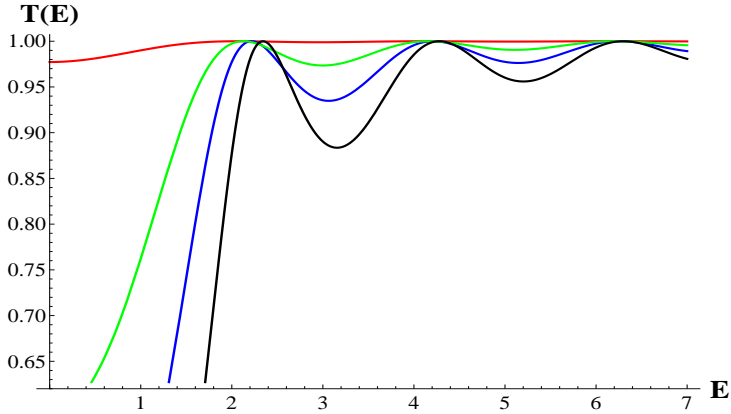}
\caption{Transimission as a function of incident energy for different
concentration of impurities (red line $E_{0}=0.1$, green line $E_{0}=0.4$,
blue line $E_{0}=1.25$, black line $E_{0}=1.7$).}
\label{transmission}
\end{figure}
In last equation, the limits of the integral are located between $\mu _{L}$
and $\mu _{L}+\left\vert e\right\vert V_{SD}$ in the case that $E_{C}>\mu
_{L}+\left\vert e\right\vert V_{SD}$. In the other side $E_{C}<$ $\mu
_{L}+\left\vert e\right\vert V_{SD}$, the upper limit will be $E_{C}$. The
conductance $G=I_{T}/V_{SD}$ can be computed up to order $E_{0}^{6}$ due to
the complexity of the integrand of last equation. The limit $\mu
_{L}\rightarrow 0$ can be taken without loss of generality, then%
\begin{equation}
G=\frac{4\left\vert e\right\vert }{\hbar V_{g}}\left(
f_{0}(V_{SD})+f_{2}(V_{SD})E_{0}^{2}+f_{4}(V_{SD})E_{0}^{4}+f_{6}(V_{SD})E_{0}^{6}+O(E_{0}^{8})\right)
\label{au20.1}
\end{equation}%
where the coefficients $f_{j}(V_{SD})$ are shown in Appendix A. In figure %
\ref{conductance}, the contributions of the different orders in $E_{0}$ can
be obtained for the conductance $G$ as a function of a dimensionless
variable $y=\frac{L\left\vert e\right\vert V_{SD}}{\hbar v_{F}}$, which is
proportional to $V_{SD}$. The limit $V_{SD}\rightarrow 0$ can be taken in $G$
and reads%
\begin{equation}
\underset{V_{SD}\rightarrow 0}{\lim }G=\frac{4\left\vert e\right\vert ^{2}}{%
\hbar }T_{0}(0)=\frac{4\left\vert e\right\vert ^{2}}{\hbar }\frac{1}{1+\sinh
^{2}(x)}  \label{au20.2}
\end{equation}%
where $x$ is a dimensionless variable which reads $x=\frac{LE_{0}}{\hbar
v_{F}}$. Last result is the correction to the minimum conductivity without
source-drain voltage. Several theoretical explanations can be found in the
literature related to the chiral nature of low energy excitations (see \cite%
{kat}, \cite{tru}, \cite{zie}). 
\begin{figure}[tbp]
\centering
\includegraphics[width=105mm,height=65mm]{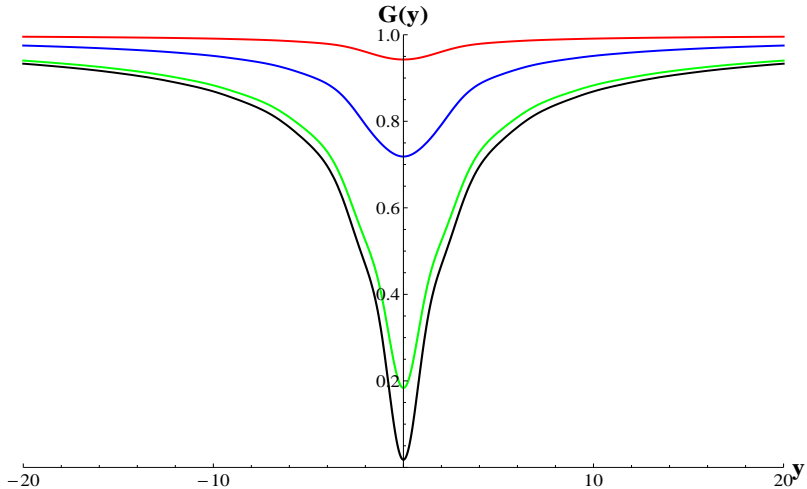}
\caption{Conductance as a function of the source-drain voltage for different
concentration of impurities (red line $E_{0}=0.2$, blue line $E_{0}=0.4$,
green line $E_{0}=0.8$, black line $E_{0}=0.84$). The unit of the $x$-axis
is dimensionless.}
\label{conductance}
\end{figure}
In figure \ref{transmission}, the electron transmission as a function of the
electron energy for different concentration of impurities is shown.\footnote{%
The energy unit of figures \ref{transmission} and \ref{length} are $eV$.} As
it is expected, the transmission probability is suppressed for energy values
less than $E_{0}$, which can be ascribed to the enhancement of the
reflection. For low impurity concentration ($E_{0}<0.2$, red line in figure %
\ref{transmission})\ transmission probability do not show transmission gap
for incident energies below $E_{0}$, similar to the results found in \cite%
{jian}. 
\begin{figure}[tbp]
\centering
\includegraphics[width=115mm,height=60mm]{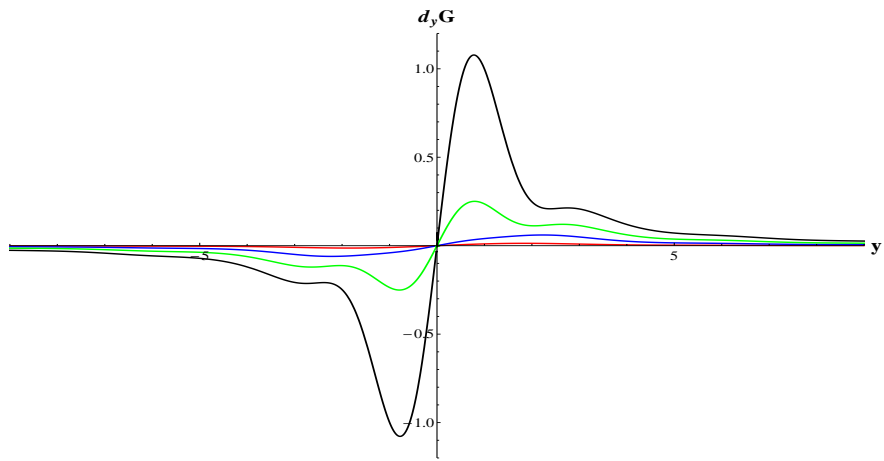}
\caption{$\frac{dG}{dy}$ as a function of the source-drain voltage (the
values of $E_{0}$ correspond to figure \protect\ref{conductance}).}
\label{schematic}
\end{figure}
In order to see what extent the transmission properties are reflected in
measurable quantities which involve averaging over the impurities in the
ballistic regime, we plot the conductance as a function of the source-drain
voltage for different concentration of impurities (see figure \ref%
{conductance}). As it is expected, the conductance decrease when the
source-drain voltage goes to zero. 
\begin{figure}[tbp]
\centering
\includegraphics[width=95mm,height=55mm]{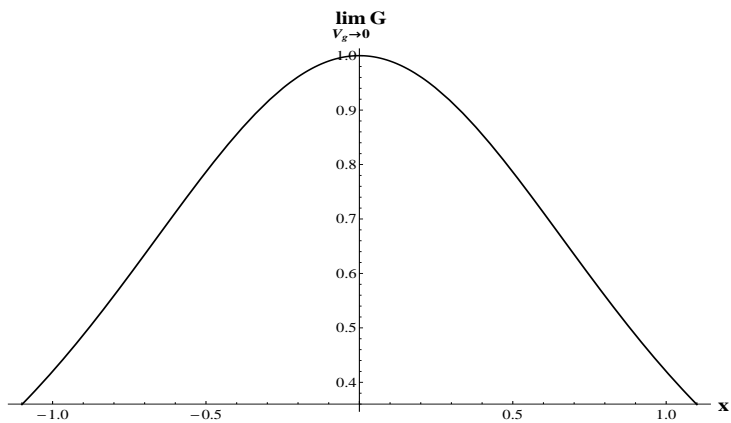}
\caption{Minimum conductance as a function of concentration of impurities.}
\label{minimum}
\end{figure}
The value of the minimum of the conductance is lower for higher values of
impurity concentrations, in corcondance with experimental reports (see \cite%
{yanbin}, fig. 2a, \cite{haijan}, fig.8, \cite{muatez}, fig. 4), where the
energy introduced by the impurities plays the rol of $kT$. In turn, the
variation of the conductance with respect the source-drain voltage can
reflect the effects found in \cite{raj} (fig. 4). The peculiar minimum value
for $V_{SD}=0$ is different than zero as it is expected for graphene. In eq.(%
\ref{au20.2}), the minimum of the conductance is computed (see figure \ref%
{minimum}). For no impurities $E_{0}=0$, the conductance is $G=\frac{%
4\left\vert e\right\vert ^{2}}{\hbar }$, although the density of states has
no charge carriers at the Fermi energy. For high values of impurities, the
minimum conductivity goes to zero, which correspond to the disentangling of
sublattices $A$ and $B$ (see eq.(\ref{tb20}) and eq.(\ref{tb21})) and no
Zitterbewegung effect.

Another relevant point is wether the effective mass $m$ of eq.(\ref{11a})
can be introduced in the Drude-Sommerfield model for diffusive conductivity.
For a simple inspection, the conductivity $\sigma $ is proportional to $%
n_{i}^{-1}$ which is a typical behavior of solids (see \cite{mahan}). In the
other side, using that the relaxation time $\tau =l/v_{F}$ and $l\sim 1/%
\sqrt{n_{i}}$ (see \cite{stauber}, \cite{jason}), the conductivity is
proportional to $\beta =n/n_{i}^{3/2}$ which is a dimensionless parameter
that separates the diffusive ($\beta <<1$) from the ballistic regime ($\beta
>>1$) (see \cite{fogler}). In this sense, the effective mass description for
the diffusive regime in graphene is highly sensitive in the relation between
charge carriers and impurity concentration.

With the purpose of obtaining a transmission coefficient with no no
dependence in the concentration of impurities, PG/DG/PG junctions can be
realized by traslating the $E_{0}$ dependence into $L$. For this, the
following renormalization equation can be obtained%
\begin{equation}
\frac{dT_{0}}{dE_{0}}=\frac{\partial T_{0}}{\partial E_{0}}+\frac{\partial
T_{0}}{\partial L}\frac{dL}{dE_{0}}=0  \label{r1}
\end{equation}%
which is a non-linear first order differential equation%
\begin{equation}
\frac{dL}{dE_{0}}=-\frac{\hbar v_{F}E^{2}}{E_{0}(E^{2}-E_{0}^{2})^{3/2}}\tan
(\frac{L\sqrt{E^{2}-E_{0}^{2}}}{\hbar v_{F}})+\frac{E_{0}L}{E^{2}-E_{0}^{2}}
\label{r2}
\end{equation}%
Transmission gap is below $E_{0}$, which implies that for practical
purposes, the r.h.s. of last equation can be expanded in Taylor series
around $E=E_{0}$%
\begin{equation}
\frac{dL}{dE_{0}}=-\frac{L}{E_{0}}-\frac{L^{3}E_{0}}{3\hbar ^{2}v_{F}^{2}}
\label{r3}
\end{equation}%
which is valid at order $O(E^{0})$. The solution of last differential
equation reads%
\begin{equation}
L(E_{0})=\pm \frac{L(1)\sqrt{3}\hbar v_{F}}{\sqrt{3\hbar
^{2}v_{F}^{2}E_{0}^{2}+2L^{2}(1)E_{0}^{2}\ln (E_{0})}}  \label{r4}
\end{equation}%
where $L(1)$ is the length of the sample of doped graphene when $E_{0}=\frac{%
V-W}{2}=1$. Last equation is valid only for incident energy $E\sim E_{0}$.
For impurity concentrations below $E_{0}=e^{-\frac{3}{2}(\frac{\hbar v_{F}}{%
L(1)})^{2}}$, real solution for $L$ does not exist. 
\begin{figure}[tbp]
\centering
\includegraphics[width=105mm,height=65mm]{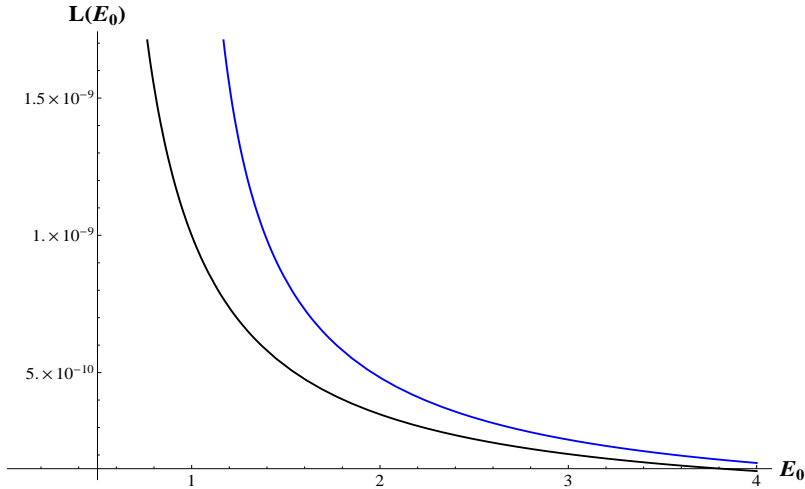}
\caption{Length of the doped graphene sample as a function of the impurity
concentration. Black line for $L(1)=10^{-9}m$ and blue line for $%
L(1)=10^{-6}m.$}
\label{length}
\end{figure}
In fig. \ref{length}, the reference value is $L(1)=10^{-9}m$ and $%
L(1)=10^{-6}m$. For values $E_{0}<1$, the length of the doped graphene
sample in both cases increase considerably. These interesting phenomena can
provide an important reference to the design of various electronic devices
based on graphene with energy gap, but where the impurities only introduces
a minimum threshold for the electron transmission coefficient. In turn, the
renormalization methods that relates the length of the sample with the
impurity concentration can give some insight of how the diffusive and
ballistic regimes are related by applying Shot noise measurements (see \cite%
{mario}), where the conductance variations are measured as a function of the
sample length of doped and clean graphene.

Finally, is interesting to note that the averaging procedure introduces a
gap in the energy band as it is occur in the Haldane model (see \cite%
{haldane} and for a more accurate version see \cite{sti}). The Haldane model
depends on an inversion symmetry breaking on-site energy $M$ for the
sublattice $A$ and $-M$ on sublattice $B$ and a complex hopping amplitude
between next-nearest neighbor due to the Peierls subtitution, which is
obtained by applying a staggered magnetic field, which is positive near the
center of each hexagon and negative near the edges which results in a zero
net flux in the hexagon. In our model, random impurities are introduced as
adsorbates that change the on-site energies where they are located. In the
model introduced in this work, the particle-hole symmetry is not destroyed
because we do not take into account next-nearest neighbor bonds. The crucial
point is that the random impurities breaks the translational symmetry and
the inversion symmetry. With the averaging procedure, the translational
symmetry is restored, but the inversion symmetry is not. In fact, the mass
term in our model depends on $V-W=\frac{1}{N}\underset{i=1}{\overset{N_{i}}{%
\sum }}(V_{i}-W_{i})$, where $V_{i}$ is the impurity located in sublattice $%
A $ and $W_{i}$ is the impurity located in sublattice $B$. If both values
are the same, $V_{i}=W_{i}$, then the mass term is zero, which implies that
Bloch electrons cannot distinguish between the upper side of the sample with
respect the other side, that is, the inversion symmetry is restored and no
gap is obtained.\footnote{%
The condition $V_{i}=W_{i}$ is not the unique condition for no gap in the
energy bands. Because we are taking an averaging over the possible locations
of the impurities, no matter if an on-site energy $V_{i}$ is not equal to $%
W_{i}$ provided that there is another on-site energy that is $W_{j}=-V_{i}$.}
In this sense, our model is a particular case of the Haldane model, where $%
t_{2}=0$ and $-W_{i}=V_{i}=M$.

\section{Conclusion}

In this paper we have studied the transport mechanism of
pristine/doped/pristine graphene junction in the ballistic regime. The
conductance as a function of the impurity concentration has been computed
using Landauer formalism with the application of the averaging procedure on
the impurity positions over the Hamiltonian. Minimum conductivity is
obtained exactly for low temperatures. A renormalization equation was
obtained for the sample length and the impurity concentration with the
purpose to obtain a transmission coefficient that do not depends on the
impurities. This result can be of importance for the manufacturing of
Schootky junctions with gapped doped graphene. Finally, the model introduced
in Section II is related to the Haldane model, finding that the band gap
obtained is due to breaking inversion symmetry introduced by the
position-averaged impurities.

\section{Appendix}

The coefficients of the Taylor expansion of the transmission of eq.(\ref{tt1}%
) as a function of $E_{0}$ up to order $O(E_{0}^{6})$ reads%
\begin{equation}
f_{0}(y)=\frac{\gamma y}{L}  \label{a1}
\end{equation}%
\begin{equation}
f_{2}(y)=\frac{L}{\gamma }(\frac{1}{y}\sin ^{2}(y)-Si(2y))  \label{a2}
\end{equation}%
\begin{eqnarray}
f_{4}(y) &=&\frac{L^{3}}{24\gamma ^{3}y^{3}}(-12y^{2}\cos
(2y)+(8y^{2}-1)\cos (4y)+1  \label{a3} \\
&&-6y\sin (2y)+2y\sin (4y)-24y^{3}Si(2y)+32y^{3}Si(4y)))  \notag
\end{eqnarray}%
and%
\begin{gather}
f_{6}(y)=\frac{L^{5}}{960\gamma ^{5}y^{5}}(12+2(3+29y^{2}-58y^{4})\cos (2y)-
\label{a4} \\
4(3+16y^{2}-128y^{4})\cos (4y)-6(1-3y^{2}+54y^{4})\cos (6y)  \notag \\
-y(33+58y^{2})\sin (2y)+16y(-3+8y^{2})\sin (4y)+(9y-54y^{3})\sin (6y)  \notag
\\
-8y^{5}(29Si(2y)-256Si(4y)+243Si(6y)))  \notag
\end{gather}%
where $\gamma =\hbar v_{F}$ and $y=\frac{L\left\vert e\right\vert V_{SD}}{%
\gamma }$, $Si(x)$ is the sine integral function.

\bigskip

\end{document}